# Exploring the Impact of Blockchain, AI, and ML on Financial Accounting Efficiency and Transformation

Vijaya Kanaparthi[1]

[1] Microsoft, Northlake, Texas, USA

**Abstract.** Continuous innovations profoundly impact the financial and commercial domains, reshaping conventional business practices. Among the disruptive forces, Artificial Intelligence (AI), Machine Learning (ML), and blockchain technology stand out prominently. This study aims to evaluate the integration of blockchain, AI, and ML within financial accounting practices. It suggests a potential revolutionary impact on financial accounting through the adoption of blockchain technology and ML, promising reduced accounting expenses, heightened precision, real-time financial reporting capabilities, and expeditious auditing processes. AI's role in automating repetitive financial accounting tasks assists organizations in circumventing the need for additional staff, thereby minimizing associated costs. Consequently, to bolster efficiency, businesses are increasingly embracing blockchain technology and AI applications in their financial accounting operations.

## 1 Introduction

Advancements in technology, particularly Artificial Intelligence (AI), have revolutionized the automation of previously complex processes [1–8]. Historically reliant on manual labor due to limited automation capabilities, the accounting and financial sector now experiences a significant transformation facilitated by AI. This transformation enables the effective execution of intricate procedures that were previously challenging. This exploration delves into the integration of AI and blockchain technology within accounting practices. While financial accounting has been gradually embracing technological advancements, its utilization has primarily focused on data archival rather than executing core accounting tasks such as transaction tracking, ledger creation, and financial statement generation. However, AI's integration is progressively reshaping financial accounting, amplifying its efficiency and impact within the field. Accounting relies heavily on technology, especially when it comes to integrating financial and management accounting [9]. The two accounting forms are linked to produce the data and information required for organizational decision-making. The application of blockchain technology to financial accounts has been aided by the growing demand for a secure business environment. Nowadays, a lot of firms engage in online trade and transactions, and both the clientele and the business place a high value on transaction security. The creation of safe online ledgers and the associated record-keeping are made possible by blockchain technology [2]. Because the transactions are encrypted using cryptography, external parties might find it difficult to intercept them.

The format of the paper is as follows: An overview of the background and context of the study is provided in Section 2. Section 3 contains a clear statement of the study's objectives. In-depth descriptions of the theoretical underpinnings and real-world applications of the suggested models are provided in Section 4. The experimental data and analysis are presented in Section 5 and offer comprehensive insights into the performance of the suggested models. Section 6 concludes by summarizing the significant discoveries, talking about their ramifications, and offering suggestions for more study.

## 2 Background

### 2.1 Blockchain

Blockchain is an accounting method that helps maintain accurate ledgers of information and deals with asset ownership changes. Blockchain technology is the most accurate and pertinent technology for meeting the technological needs of the accounting profession since financial accounting requires precise and reliable information. Ledgers with superior qualities like transparency, security, accountability, permanence, and immutability are made possible by blockchain technology [10]. Due to its decentralized nature, the system is completely safe and effective for transactions. Blockchain technology can create immutable data records through a quick and accurate automated procedure. Blockchain, being a distributed ledger technology, has a big impact on financial accounting and auditing. There are numerous applications for blockchain technology in financial statement audits. First, because it continuously records the principal ledgers, it offers data for evidence. Since the ledger and financial records stored by the technology are permanent, they can be recovered at a later time. Additionally, they save the auditors' and accountants' time that could have been spent organizing and examining the ledgers. Building safe, transparent, and effective smart contracts is made feasible by blockchain technology. By offering a preliminary examination of the information, blockchain technology is utilized in banks and other financial institutions to identify individuals and track their performance patterns [11]. The most significant advantage of blockchain technology is its ability to establish a connection between the confidentiality of user information, technology, or transactions and the openness of business process audits. Since there are typically no middlemen involved in the transactions, intelligent contracts have lower transaction costs. The contract's transactions are also carried out in real time and recorded in digital ledgers that are open to all network users. The primary goal of technology is to decrease errors in record-keeping and commercial transactions. Blockchain technology changes the accounting process from a traditional double-entry system to a triple-entry system, which records transactions and reduces errors to almost nothing. Reducing errors in financial records improves their dependability and usability because the forms can help the organization make decisions and prepare appropriately.

## 2.2 Artificial Intelligence

The accounting industry is undergoing a transformation thanks to AI, which is automating tedious operations and boosting data management and accuracy. Many broken processes that have long defied technology developments afflict the accounting cycle. Although accounting operations have not been fully integrated, the sheer volume of repetitive tasks has increased demand for a system that can lessen that load. The most sought-after answer to this issue is AI, which also has the added advantages of cloud-based data storage, mistake reduction, and simplified reconciliation. The value of financial accounting records is determined by their accessibility and accuracy. AI ensures the authenticity of these records by maintaining their up-to-date status and securely storing them via cloud-based systems. The landscape of financial technology has fundamentally altered business operations, swiftly supplanting traditional cash transactions with digital counterparts. Absent AI's assistance, the capture and monitoring of transactions carried out through credit cards, mobile payments, vouchers, and various electronic mediums would pose significant challenges. Businesses leverage AI to preserve sales data and embrace modern payment modalities, establishing a robust framework for streamlined financial accounting. Furthermore, platforms rooted in blockchain for sales and AI-driven systems such as QuickBooks have notably hastened this transformation.

## 3 Objectives

This study focuses on evaluating the utilization of blockchain technology and AI within the realm of financial accounting. The primary aim is to illustrate the advancements made in financial accounting due to AI integration, identifying the factors that promote the fusion of AI and blockchain technology in accounting practices. Additionally, it seeks to outline actionable steps to enhance the efficiency of AI and blockchain integration in financial accounting processes. Furthermore, this paper intends to delve into the challenges and limitations inherent in employing AI and blockchain technology in financial accounting. It will explore the potential impact of these novel technologies on the entire accounting sector and the broader financial landscape. The insights gained from this study will serve as valuable guidance for software developers, lawmakers, and accounting professionals, providing a comprehensive understanding of the evolving landscape of financial accounting in the digital era.

## 4 Methodology

This study investigates the impact of blockchain technology, Machine Learning (ML), and Artificial Intelligence (AI) on the landscape of financial accounting by conducting an extensive analysis of existing literature. The secondary data incorporates primary research exploring the utilization of blockchain technology, AI, and ML within the realm of finance, particularly focusing on their role in financial accounting. The results will encompass a comprehensive evaluation of the research findings, accompanied by an assessment of their alignment with the objectives outlined in the paper. A portion of the dataset used in this investigation is included in Table 1. For the chosen papers, there were roughly 185 studies in the dataset overall. The studies had to be written using primary data, and the research had to discuss the applications of blockchain, AI, and ML in financial accounting.

**Table 1.** Overall Studies

| Journal | Number of studies | Keywords |
| --- | --- | --- |
| Review of Finance | 1 | Blockchain, corporate governance |
| Review of Financial Studies | 3 | Blockchain, FinTech, innovation |
| British Accounting Review | 2 | Cloud, big data, blockchain, and AI |
| Information and Organization | 2 | Digital innovation and transformation, blockchain |
| Journal of Financial Reporting and Accounting, | 2 | Blockchain, accounting, Financial Accounting |
| Journal of Economics and Finance | 2 | Financial Accounting AI |
| Procedia Computer Science | 2 | Fraud Detection, ML, AI |

## 5 Results

The numerous research that have been examined demonstrate the use of blockchain, AI, and ML in financial accounting. To demonstrate how the three aforementioned elements have impacted financial accounting, this section examines the studies mentioned above.

### 5.1 Blockchain Technology in Financial Accounting

Blockchain technology is a noteworthy discovery that has the potential to completely transform a number of industries, including accounting for financial services. Decentralization, immutability, and transparency—three fundamental features of blockchain technology—make it an effective instrument for the financial accounting industry. The financial infrastructure plays a critical role in decision-making, performance analysis, control mechanism development, and employee incentive design. An essential component of financial accounting are ledgers. Managing ledgers is one of the initial tasks in financial accounting. According to Zachariadis et al. [12], blockchain technology is disruptive in 2019 because it provides numerous opportunities for ledger management that are not achievable with current accounting technologies. Although there have been obstacles to blockchain adoption in many nations, the technology's numerous applications in financial accounting have encouraged its implementation. The applications of blockchain technology in financial accounting are compiled in Table 2 below.

**Table 2.** Applications Of Blockchain Technology

| Use | Explanation |
| --- | --- |
| Asset tracking | Blockchain is used to track the ownership and transfer of assets such as real estate, stocks, and bonds using a distributed ledger system |
| Audit trails | All data on a blockchain is tamper-proof and can be traced back to its origin, making it easier to detect and prevent fraudulent activity. |
| Smart contracts | Smart contracts are used to automate complex financial transactions, such as issuing bonds or settling derivatives. |
| Identity verification | Blockchain provides a secure and decentralized method of verifying the identities of people, which can be used for Know Your Customer (KYC) and Anti-Money Laundering (AML) compliance in financial accounting. |
| Payment processing | Blockchain technology provides faster, cheaper, and more secure payment processing by eliminating intermediaries and reducing transaction costs. It also facilitates cross-border payments by reducing the need for currency conversions. |
| Supply chain management | Blockchain is used to track the movement of goods and materials throughout the supply chain hence increasing transparency, reducing fraud, and improving efficiency. |
| Regulatory compliance | Blockchain helps to ensure compliance with various financial regulations, such as data privacy laws, tax regulations, and securities |

laws. By providing a tamper-proof and auditable record of transactions, blockchain makes it easier to comply with regulatory requirements.

By eliminating a single point of failure or control, decentralizing the ledger system lowers the possibility of fraud and manipulation. Financial accounting benefits greatly from blockchain technology's distributed ledger system characteristic, which makes a distributed ledger resistant to unauthorized alterations. As a result, stakeholders can have more faith in the system because the integrity of financial records is preserved. Their usefulness is greatly influenced by financial visibility and the capacity to guarantee the integrity and correctness of records. Other important aspects of blockchain technology include the immutability and permanence of records. When tracking transactions and even performing forensic auditing, a system's capacity to guarantee that a transaction cannot be changed or withdrawn from the ledger once it has been recorded is crucial [13]. Due to its ability to produce a trustworthy and auditable transaction trail, this feature is essential for financial accounting. Immutability lowers the possibility of inconsistencies and errors by allowing auditors and regulators to access an accurate historical record. Auditors are always required to categorize transactions using a variety of techniques in order to highlight fraudulent activity [14–17]. But when blockchain technology is used, the system creates an audit trail, which improves the effectiveness and efficiency of the editing process. A double-entry system is used in financial accounting to guarantee the accuracy and transparency of financial information. Since it makes the full transaction history visible to all network users, transparency is a crucial component of blockchain technology. This degree of transparency in financial accounting improves the dependability and accuracy of financial records, making it simpler for auditors to confirm transactions and spot any anomalies. A triple-entry system can further improve the procedure and offer a clear, accurate, and trustworthy record thanks to blockchain technology. Examining the most effective ways to enhance the financial accounting system has drawn a lot of interest. In order to improve the quality and dependability of financial data, [18] looked into how blockchain could change the accounting process by switching from a double-entry to a triple-entry method. The authors also emphasized how blockchain technology might be advantageous for auditing, since it could offer real-time access to financial data and lower the possibility of fraud. A potential answer to some of the relevant problems with financial accounting records' correctness is blockchain technology. To make business blockchain applications a reality in the financial accounting industry, a few major obstacles need to be overcome. Scalability, interoperability, data protection, and regulatory compliance are some of these difficulties. The author proposed that the full potential of blockchain technology in financial accounting may be unlocked by addressing these issues through cooperation and standardization initiatives. Another significant use of blockchain technology in financial accounting is smart contracts. When the pre-established requirements are satisfied, these self-executing contracts automatically enforce the terms and conditions of a transaction. Financial transactions are completed more quickly and efficiently thanks to this method, which also decreases transaction costs and eliminates the need for middlemen. The possible advantages and difficulties of using blockchain technology into financial accounting systems were investigated by [10]. They emphasized how blockchain might improve financial reporting's efficiency and transparency by giving regulators, auditors, and other stakeholders instant access to financial data. The authors did, however, also note the regulatory and technological difficulties that come with integrating blockchain technology into financial accounting, such as the requirement for suitable infrastructure and adherence to current laws.

To sum up, there is a lot of potential for the financial accounting industry using blockchain technology. Smart contracts can increase the efficiency of financial transactions, while its fundamental characteristics—decentralization, immutability, and transparency—can improve the accuracy and dependability of financial records. Blockchain technology has the ability to completely change financial accounting and auditing procedures, making them more transparent, safe, and effective—even though there are still issues to be resolved.

### 5.2 Artificial Intelligence in Financial Accounting

Another revolutionary technology that is changing several sectors, including financial accounting, is AI. AI systems are capable of making judgments with little assistance from humans, learning from data, and adapting to novel circumstances. The efficiency, accuracy, and efficacy of accounting procedures can all be markedly increased by using AI in financial accounting [19]. Data input, reconciliation, and report production are examples of repetitive operations that AI is mostly used for in financial accounting. AI-powered software reduces error risk and frees up accountants' time to work on more important responsibilities by processing massive volumes of data faster and more correctly than human accountants. Table III shows the use of AI in the field of financial accounting.

**Table 3.** AI in Financial Accounting

| Function | Explanation |
| --- | --- |
| Automating data entry | AI helps financial accountants automate data entry tasks by recognizing and extracting information from invoices, receipts, and other financial documents. This can reduce errors and save time. |
| Fraud detection | AI is used to identify patterns and anomalies in financial data that may indicate fraud or other irregularities. |
| Financial forecasting | AI helps in the analysis of financial data to help predict future trends and outcomes, which can assist in strategic |
|                              | decision-making.                                                                                             |
| Personalized financial advice | AI-powered chatbots provide personalized financial advice to customers based on their financial situation and goals. |
| Risk management              | AI assists financial accountants identify and assess risks associated with investments, loans, and other financial transactions. |

Enhancing financial analysis and decision-making stands as a crucial application of AI within financial accounting. AI-driven systems adeptly handle extensive datasets, uncovering trends, patterns, and irregularities that might elude human accountants. This translates into substantial long-term benefits for businesses, empowering them with improved financial planning, risk management, and wiser investment decisions. Moreover, AI plays a pivotal role in the realm of financial accounting by detecting and preventing fraud. Real-time fraud detection becomes pivotal in halting deceitful activities and mitigating the impact of any fraudulent occurrences. AI-powered solutions extend their reach to aiding tax compliance and planning. By delving into intricate tax laws, these AI technologies pinpoint potential avenues for tax savings, enabling individuals and corporations to optimize their tax strategies. Furthermore, AI becomes a valuable ally in identifying tax fraud. Leveraging machine learning algorithms, it flags atypical transactions or trends that could signify attempts at tax evasion or avoidance. As AI streamlines repetitive tasks, augments financial analysis, and fortifies fraud detection, its potential impact on financial accounting becomes substantial. Successful implementation of AI promises heightened efficiency, precision, and efficacy in accounting operations, notwithstanding the hurdles like data quality, integration, and legislative challenges that need overcoming. Through AI integration, accounting professionals can elevate their roles, focusing on more critical tasks, ultimately steering companies toward improved financial outcomes.

## 5.3 Artificial Intelligence and Blockchain Integration in Financial Accounting

The convergence of Artificial Intelligence (AI) and blockchain technology holds substantial promise within the realm of financial accounting. AI's adeptness in data analysis and automation synergizes with blockchain's capacity to generate secure, transparent, and immutable records. By integrating these technologies, organizations stand to enhance the precision, efficiency, and safeguarding of their financial accounting processes [20]. The potential impact of this amalgamation on financial accounting is profound. However, realizing its complete potential necessitates surmounting certain hurdles. These challenges encompass establishing definitive guidelines for the application of these technologies in financial accounting, addressing the scalability of blockchain networks, and seamlessly integrating AI systems with existing accounting protocols and systems. As ongoing research and development progress in blockchain and AI, resolving these hurdles will pave the way for a new era in financial accounting, propelled by cutting-edge technology.

### 5.4 Machine Learning in Financial Accounting

Machine Learning (ML) plays a significant role in the realm of financial accounting, leveraging features akin to blockchain technology and AI. Its application within the financial sector spans a wide spectrum. Within Table 4, a comprehensive summary of several prominent applications is delineated. AI and ML are indispensable in discerning transactional patterns, especially in the detection of fraudulent activities and potential scams within a company.

**Table 4.** ML in Financial Accounting

| Function | Explanation |
| --- | --- |
| Predictive modeling | ML is used to build predictive models that analyze financial data to help forecast future trends and outcomes. |
| Fraud detection | ML is important in identifying patterns and anomalies in financial data that may indicate fraud or other irregularities. |
| Risk management | ML identifies and assesses risks associated with investments, loans, and other financial transactions. |
| Automating data entry | ML is used to automate data entry tasks by recognizing and extracting information from invoices, receipts, and other financial documents. |
| Personalized financial advice | ML-powered chatbots provide personalized financial advice to customers based on their financial situations and goals. |
| Credit scoring | ML is used to analyze credit history and other financial data to help determine creditworthiness and calculate credit scores. |
| Compliance monitoring | ML is used to monitor financial transactions for compliance with regulations and policies. |

## 5 Analysis

In this segment, an evaluation is presented concerning the discoveries in connection with contemporary global data. This portion specifically scrutinizes the utilization of blockchain technology in accounting and the consequent financial benefits derived from the amalgamation of blockchain and AI. The visual representation of the survey outcomes can be observed in Fig. 1.

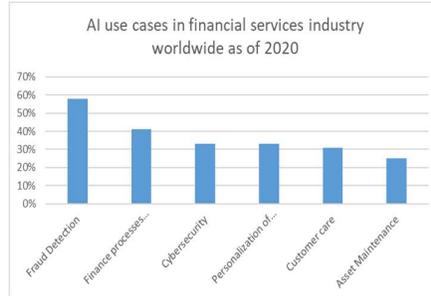

**Fig. 1.** AI use in the financial industry

The conclusions drawn from these findings highlight that AI predominantly serves the realm of financial accounting, particularly in the areas of fraud detection and financial processes [21]. Other applications of AI that have been found include asset maintenance, customer service management, cybersecurity, and process personalization. In order to identify irregularities like fraud within an organization, AI must be able to learn algorithms and map patterns in transactions. Understanding the AI index and its most popular applications in the modern world is crucial to comprehending how ML and AI enhance financial accounting. The indices of the various AI applications are summarized in Fig. 2.

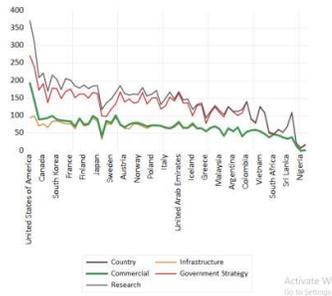

**Fig. 2.** Index of AI use scores

The Fig. 2 shows that the use of AI has the highest score for government strategy, infrastructure development, research, and commercial uses. The main concern of this paper is to investigate the commercial uses of AI in financial accounting. It is expected that the usage of AI in accounting will continue to increase in the near future. The market for the use of AI in accounting is segmented by component, deployment mode, technology, enterprise, application, and region. The Fig. 3 shows the different segmentations of AI in accounting.

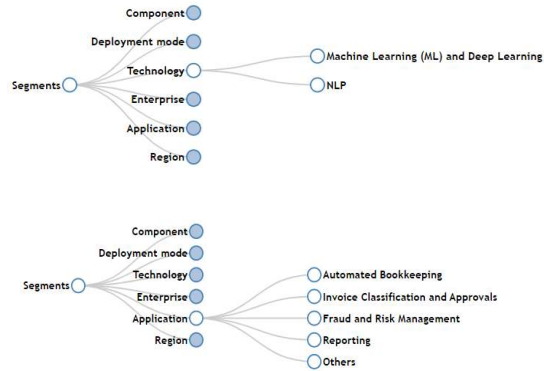

**Fig. 3.** Segmentation of AI in accounting

AI is becoming more and more in demand in financial accounting. The projected valuation of AI in the accounting industry by 2030 is displayed in the Fig. 4.

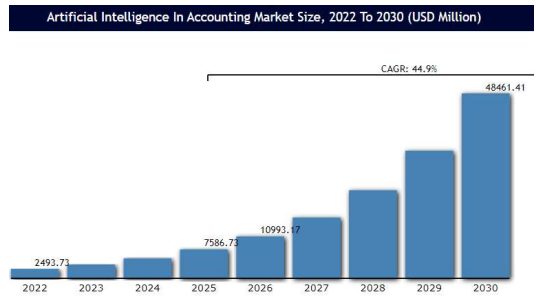

**Fig. 4.** Projected valuation of AI in accounting

According to the Fig. 4, the financial accounting AI market is expected to grow in value by a constant percentage until 2030. ML services that interpret accounting data and support decision-making inside an organization are included in AI. The majority of firms easily integrate AI into their services since they believe it to be more productive and cost-effective. AI contributes to lower financial accounting and auditing expenses. By lowering the cost of ledger production and reconciliation and by introducing uncertainty into asset ownership, blockchain contributes to the reduction of financial accounting process expenses. This makes it essential to verification and audit procedures. In the financial industry, blockchain has gained a lot of attraction because of the aforementioned services.

The Fig. 5 illustrates how the number of blockchain users has increased over the previous six years.

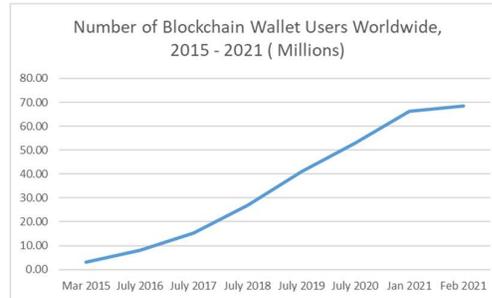

**Fig. 5.** Number of blockchain users

Blockchain has the benefit of being applicable to all facets of accounting, including financial accounting. According to a Deloitte blockchain survey, the use of blockchain is depicted in Fig. 6.

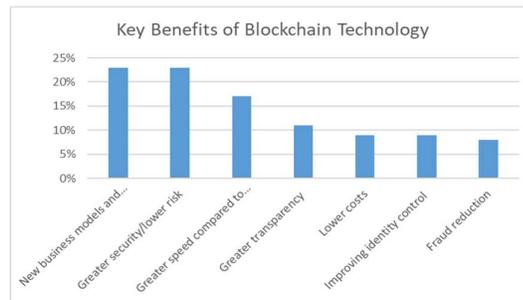

**Fig. 6.** Benefits of Blockchain

The Fig. 6 makes it clear that the majority of blockchain users think that new business models, enhanced transaction security, and lower risks are the three main advantages of blockchain. Increased speed in financial accounting, increased transparency, decreased costs, and decreased fraud are some of the other major advantages of blockchain technology in finance. The Fig. 7 illustrates how blockchain is being used around the world.

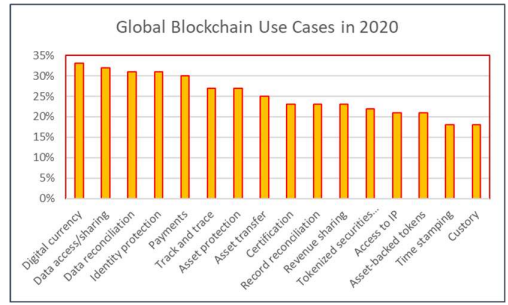

**Fig. 7.** Global uses of blockchain technology

The Fig. 7 indicates that digital currency, data sharing, and data reconciliation are the three areas where blockchain technology is most frequently used. The Fig. 7 demonstrates how helpful blockchain technology is for financial accounting procedures like revenue management, record reconciliations, tracking transactions, and payment management. Various industries make investments in technologies that they believe are essential to the services they provide. The Fig. 8 illustrates how much money various sectors have invested in blockchain technology.

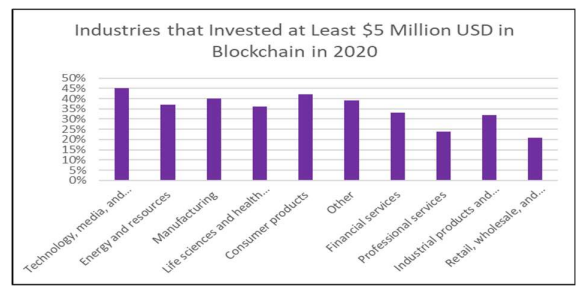

**Fig. 8.** Investment of different industries in blockchain technology

Financial services are one of the largest industries with over $5 million invested in 2020, as the Fig. 8 illustrates. The sector employs blockchain technology extensively, which explains why there is a significant readiness to invest in it. According to research, the use of blockchain in financial accounting has forced auditing practices to change in order to comply with current requirements. Blockchain technology lowers accounting costs by doing away with the need for middlemen, such banks or other financial organizations, to verify and process financial transactions. Transaction fees are levied by intermediaries, which raises the expenses an organization pays for transactions. When accountants reconcile financial figures, such as the bank balance between the cash book and the real

amount in the bank, they incur additional fees that further complicate the process of financial accounting. Transactions may be handled swiftly and effectively with blockchain technology instead of requiring costly and time-consuming manual reconciliation [22, 23]. It has been demonstrated that ML is essential to blockchain technology and AI. Many of the duties that accountants have historically performed in an organization have been automated by ML in the financial sector. ML algorithms, for example, are used to analyze financial data and spot patterns and anomalies that could point to the possibility of fraudulent or incorrect transactions. ML is also used to optimize financial processes and enhance decision-making, reduce fraud, and make repetitive accounting tasks easier to complete. One application of ML is the extraction of information from invoices for payments and accounting, particularly when the invoices have a predetermined format and can be extracted by a program that has been trained to do so. The aforementioned instance primarily applies in Italy, where all bills must be electronic and adhere to a specific format per legislation. The preparation of financial statements and documentation, as well as the auditing of the financial accounts, are made simple by the application of ML. Additionally, ML algorithms are employed in the analysis of past financial data to spot trends or patterns that are crucial for finance professionals to recognize in order to improve forecasts, investment strategies, and budgeting. Within a corporate setting, access to reliable financial information stands as a linchpin for informed decision-making. Machine Learning (ML) integrated into automated data processing significantly aids in steering pivotal decisions about an entity's projected trajectory. Choices in finance reverberate across a company's entire framework, impacting its future trajectory. The synergy between Blockchain technology and ML dynamically supports the expansion of the financial landscape. Real-time financial reporting, enabled by Blockchain and ML, offers a dual advantage: reducing financial accounting costs while augmenting precision. Blockchain facilitates the real-time collection and updating of financial data, furnishing up-to-the-minute insights into an organization's financial well-being. ML algorithms applied to this data yield analytical results disseminated to relevant stakeholders. Not only does real-time reporting save the laborious pursuit of individual records, but it also ensures adherence to accounting standards. The time stamps embedded within transactions prove invaluable for tracking and validating documentation. In essence, Blockchain technology serves as an incorruptible ledger, encapsulating a comprehensive log of all corporate transactions. Its implementation ensures a meticulous record of transactions, uniquely linked to system users, effectively eliminating the potential for fraud or errors inherent in manual financial data recording or centralized database storage. This automation of data verification and validation through Blockchain significantly reduces errors, enhancing the accuracy of financial reports. The cumulative impact of Blockchain in preserving transaction integrity instills confidence in organizations regarding the security of their assets by eradicating fraudulent practices. Research endeavors have been fervently focusing on harnessing Blockchain technology, AI, and ML to refine organizational auditing processes. The transparent, verifiable, and dependable ledger system offered by Blockchain greatly empowers auditors to ascertain the reliability of financial data and detect potential discrepancies within financial statements. The foundation of robust internal controls necessitates evidence for data comparison, and in this aspect, a Blockchain's immutable nature renders it exceptionally trustworthy. Moreover, the integration of AI and ML enables the automation of select accounting operations, streamlining processes within the financial realm.As an illustration, machine learning (ML) and artificial intelligence (AI) play a significant role in streamlining audits by automatically extracting details from receipts and invoices, sparing auditors the manual effort. In essence, this investigation focuses on the impact of blockchain technology, AI, and ML on financial accounting. According to the analysis, financial accounting encompasses the meticulous recording of financial data and transactions, consolidating them into financial statements, and leveraging these statements to document an entity's business transactions. The research suggests that blockchain technology and ML have the potential to transform financial accounting practices. They could effectively reduce accounting expenses, elevate precision, enable real-time financial reporting, and expedite the audit process. Through the automation of repetitive financial accounting tasks, AI assists organizations in evading additional expenses associated with hiring more personnel for accounting purposes. Therefore, it is recommended that enterprises adopt these technologies to augment the efficiency of their financial accounting framework.

We emphasize the use of figures (Fig. 1 to Fig. 8) and other visual aids to communicate survey results and trends. The figures display information about the application of AI, how it is segmented in the accounting sector, how much the sector is expected to pay for AI, how many people use blockchain technology globally, how much money is invested in blockchain technology across different industries, and how the market for AI in financial accounting is expected to grow until 2030.

### 6.1   Security Risk

The technologies of ML, blockchain, and AI in financial accounting raise a number of security issues. Risks associated with blockchain include smart contract vulnerabilities and consensus procedures, as well as the possibility of 51% attacks jeopardizing network stability. Due to their reliance on large datasets, AI systems must adhere to strict data privacy regulations in order to prevent unwanted access. The security challenges in AI are further intensified by adversarial assaults and model bias. Problems with machine learning include model inversion, data poisoning, and the possibility of overfitting or underfitting during training. Concerns about data exchange security, interoperability, and the human component are brought up by the integration of these technologies, underscoring the significance of strong user authentication and preventing insider threats. Regulatory compliance introduces an extra layer of security issues, especially with regard to data protection and following legal requirements. To meet these problems and guarantee the safe integration of new technologies into financial accounting systems, a thorough security plan that includes encryption, ongoing monitoring, regulatory compliance, and frequent security audits is required.

However, important factors including the robustness of the consensus process, smart contract code audits, data privacy safeguards, encryption standards, resilience to adversarial assaults, and regulatory compliance must be taken into consideration when assessing the security of blockchain, AI, and ML in financial accounting. Other important factors would include the defense against insider threats, the efficacy of user authentication techniques, and the general robustness of integrated systems.

## 7 Conclusion and Future Directions

The landscape of financial accounting has experienced a significant evolution propelled by the integration of blockchain, AI, and ML innovations. These technologies have notably bolstered security measures, operational efficiency, and the overall transparency of accounting processes. They've specifically elevated functionalities like data input, risk assessment, and the identification of fraudulent activities. While challenges around scalability and regulatory compliance persist, the fusion of these advancements holds immense promise in reshaping the realm of financial accounting. Future work should focus on improving blockchain scalability, creating uniform rules, protecting privacy and securityand carrying out more practical application tests of these technologies in financial accounting. These programmes will enhance and optimise the benefits of this unique pairing.